\documentclass[aps,prl,twocolumn,groupedaddress,showpacs,floatfix]{revtex4}
\usepackage{graphicx}
\usepackage{subfigure}
\usepackage{amsmath}
\def\ket#1{|#1\rangle}

\newcommand{\fref}[1]{Fig.~\ref{#1}}

\begin{document}
	
\author{N. Arunkumar$^{1}$, A. Jagannathan$^{1,2}$, and J. E. Thomas$^{1}$}
\affiliation{$^{1}$Department of  Physics, North Carolina State University, Raleigh, NC 27695}
\affiliation{$^{2}$Department of Physics, Duke University, Durham, NC 27708}
\date{\today}

\title{Probing Energy-Dependent Feshbach Resonances by Optical Control}

\begin{abstract}

Optical control enables new high resolution probes of narrow collisional (Feshbach) resonances, which are strongly dependent on the relative momentum of colliding atom pairs, and important for simulating  neutron matter with ultracold atomic gases.   We demonstrate a two-field optical vernier, which expands kHz (mG) magnetic field detunings near a narrow resonance into MHz optical field detunings, enabling precise control and characterization of the momentum-dependent scattering amplitude. Two-photon loss spectra are measured for the narrow resonance in $^6$Li, revealing rich structure in very good agreement with our theoretical model. However, anomalous overall frequency shifts between the measured and predicted two-photon spectra are not yet explained.


\end{abstract}

\maketitle

Magnetic Feshbach resonances~\cite{ChinFeshbach} in ultracold gases have been extensively exploited to achieve important milestones in atomic physics, from the realization of strongly interacting Fermi systems~\cite{OHaraScience,ZwierleinFermiReview,ZwergerReview} to the observation of  Efimov trimers~\cite{RGrimmEfimov}. Typically, in a magnetic Feshbach resonance, an external magnetic field is used to tune the total energy of two colliding atoms in an energetically open channel into resonance with a bound dimer state in a closed channel. For a narrow Feshbach resonance, where the width is comparable to the relative energy of the incoming atom pair, the interactions are strongly momentum-dependent~\cite{HoNarrowFB,OHaraNarrowFB}. Momentum-dependent narrow resonances offers important possibilities for realizing novel quantum phases in ultracold gases, such as synthetic FFLO pairing states~\cite{OptcontrolCOM, FuldeFerrell}, where resonant interactions occur only for nonzero center of mass momentum, and breached-pair superfluids~\cite{BreachedPair}, where superfluid and a normal components are accommodated in different regions of momentum space.

The large effective range $r_e$ of narrow Feshbach resonances,  coupled with resonant interactions, can be exploited to simulate neutron matter at sub-nuclear densities, where $k_F\,r_e \gtrsim 1$, with $k_F$ the Fermi momentum. This regime is important for understanding the physics of neutron stars and supernova~\cite{PethickNarrow}. For the narrow Feshbach resonance in $^6$Li near 543.27 G  (width $\Delta\,B \simeq 0.1 $G) $r_e$ is anomalously large, $r_e \approx - 7 \times 10^4\,a_0$, with $a_0$ the Bohr radius~\cite{OHaraNarrowFB,WuThomasPRAOptControl}. A fundamental understanding of the momentum structure of narrow Feshbach resonances is therefore of great interest to both the atomic and nuclear physics communities.

Previous experimental studies of narrow Feshbach resonances with momentum-dependent interactions have employed scans of external magnetic fields to create narrow Feshbach molecules~\cite{RandyNarrow}, to measure two-body interactions~\cite{OHaraNarrowFB}, and to study three-body recombination loss~\cite{LuoNarrow}. However, these studies have gained limited insights into the momentum-dependence of interactions in a two-body scattering process. This is partly due to the limited momentum resolution achieved in employing a magnetic field scan near a Feshbach resonance, where, for example, with $^6$Li atoms, a thermal energy for $1 \mu K $ = h x 20 kHz is equivalent to 7 mG of B-field tuning at 2.8 MHz/G.
Manipulation of Feshbach resonances also has been accomplished by using optical fields to tune the closed channel molecular bound state across the open channel continuum~\cite{PLettOFB,EnomotoOFB,TheisOFB,YamazakiSpatialMod,RempeOptControl,CetinaOptcontrolPRL,ChinMagicOptControl}. However, optical methods have had limited applicability due to atom loss arising from spontaneous scattering, which limits the tuning range.

Recently, we demonstrated optical control of two-body interactions in $^6$Li using closed-channel electromagnetically induced transparency (EIT) to suppress optical scattering and subsequent loss~\cite{ArunOptcontrol}. In this scheme, two optical fields are used to tune the closed channel molecular bound state near a magnetic Feshbach resonance with suppressed atom loss through destructive quantum interference~\cite{WuOptControl1, WuOptControl2}.

In this Letter, we demonstrate that closed-channel EIT provides a high-momentum-resolution probe of the two body scattering amplitude, which determines the momentum dependence for both elastic and inelastic scattering.  We measure loss spectra near a narrow Feshbach resonance in $^6$Li as a function of two photon-detuning and observe widely different spectra for magnetic fields on the atomic (BCS) side above the resonance compared to the molecular (BEC) side below the resonance. Near the minimum loss point at the two-photon resonance, the spectra reveal a rich structure, which is extremely sensitive  to the momentum dependence of the scattering amplitude.  The spectra predicted by our momentum-dependent continuum-dressed state model are in excellent agreement with the data, both in shape and in absolute magnitude. However, we observe unexplained overall frequency shifts between the measured and predicted two-photon spectra, which are nearly independent of magnetic fields below resonance and strongly dependent on magnetic field above resonance.

\begin{figure}[htb]
	\centering
	\includegraphics[width = 3.4 in]{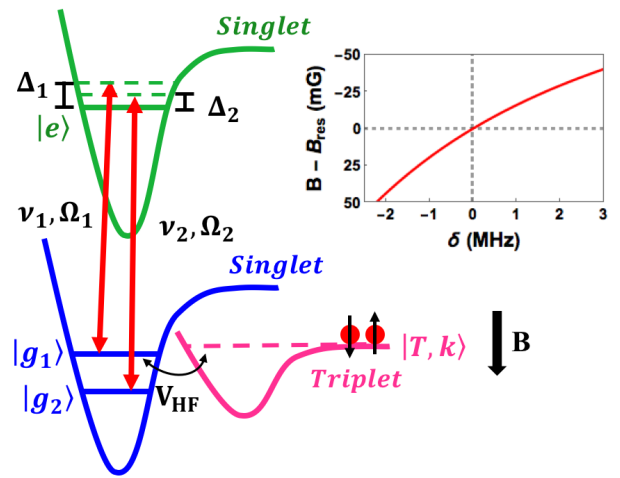}
	\caption{Level scheme for closed-channel electromagnetically induced transparency (EIT). Optical fields $\nu_1$ and $\nu_2$ couple the ground molecular singlet states $\ket{g_1}$ and $\ket{g_2}$ with the excited singlet state $\ket{e}$ of the closed channel, resulting in a light shift of state $\ket{g_1}$. Atoms reside in the open channel triplet continuum $\ket{T,k}$, which is hyperfine coupled to $\ket{g_1}$, producing the narrow Feshbach resonance. Inset: The red line shows the effective magnetic tuning $B-B_{res}$, where $B_{res} = 543.27$G, as a function of two-photon detuning $\delta$ (for $k=0$). The horizontal dashed line, $B-B_{res} = 0$, corresponds to the unshifted position of $\ket{g_1}$ for $\delta =0$. The energy of $\ket{g_1}$ tunes almost linearly with $\delta$ near the unshifted position, corresponding to a magnetic field tuning of $\simeq -18 {\rm mG}\times\delta$(MHz).\label{Horizontalimage}}
\end{figure}

The basic level scheme for our closed-channel EIT method is shown in \fref{Horizontalimage}. Optical fields  $\nu_1$ and $\nu_2$, with detunings $\Delta_1$ and $\Delta_2$ and  Rabi frequencies $\Omega_1$ and $\Omega_2$, couple the ground molecular states of the singlet potential, $\ket{g_1}$ and $\ket{g_2}$, to the excited state $\ket{e}$. A narrow Feshbach resonance arises from the second order hyperfine coupling $V_{HF}$ between the bound state $\ket{g_1}$ and the triplet continuum $\ket{T,k}$, which tunes downward with increasing magnetic field $B$. We initially choose a magnetic field such that the triplet continuum is tuned near $\ket{g_1}$, i.e.,  close to the resonance magnetic field $B_{res} = 543.27$~G~\cite{OHaraNarrowFB}. Near resonance, the optical detunings $\Delta_1$ and $\Delta_2$ are large compared to the magnetic detunings $\frac{2\,\mu_B}{\hbar}\,(B-B_{res})$, so that the two-photon detuning $\delta \simeq \Delta_2-\Delta_1$. For $\delta =0$, the light shift of $\ket{g_1}$ vanishes.  As $\delta$ is varied from negative to positive, $\ket{g_1}$ tunes upward, from below to above its unshifted position. From Eq.~\ref{eq:1} below, we find that tuning $\delta$ is equivalent to magnetically tuning $\ket{T,k}$ with a magnetic field of \mbox{$\simeq -\frac{|\Omega_1|^2}{|\Omega_2|^2}\frac{\hbar\,\delta}{2\mu_B}\simeq -18\,{\rm mG}\times\delta$(MHz)} for our optical parameters. Hence,  $\delta$ acts as an optical vernier to investigate and control the fine momentum-dependent features of the narrow Feshbach resonance, where the magnification in resolution scales as $ \frac{|\Omega_1|^2}{|\Omega_2|^2}$ . We note that the required stability of the chosen magnetic field is the same as for magnetic tuning, but the need for ultra-high resolution magnetic field scans is circumvented.

We begin with a cloud of $^6$Li atoms, confined in a CO$_2$ laser trap, and prepared in a 50-50 mixture of the two lowest hyperfine states, $\ket{1}$ and $\ket{2}$. We evaporatively cool the atoms at 300 G to a typical temperature $T\simeq 2.0\,\mu$K~\cite{Range} in the nondegenerate regime, to clearly display the momentum dependence of the loss. We ramp the magnetic field to 528 G, near the zero crossing in the scattering length, and apply an RF sweep (30 ms) to transfer the atoms from state $\ket{2}$ to state $\ket{3}$.  The magnetic field is then ramped to the field of interest, where we wait 3 s for the magnetic field to stabilize. We then apply the $\nu_2$ beam with a typical Rabi frequency $\Omega_2 = 2\pi\times 26.0$ MHz~\cite{Range}.  We wait 30 ms for the atoms to reach equilibrium in the combined potential of the CO$_2$ laser and $\nu_2$ beams. An RF $\pi$ pulse (1.2 ms) then transfers the atoms from state $\ket{3}$ to state $\ket{2}$. The $\nu_1$ beam with Rabi frequency $\Omega_1 = 2\pi\times 5.9$ MHz~\cite{Range} and detuning $\Delta_1 = +2\pi\times 19$ MHz is then applied for 5 ms. The atoms are imaged after a time of flight of 250 $\mu$s, to determine the total atom number.

\begin{figure}[htb]
	\centering
	\includegraphics[width = 3.4 in]{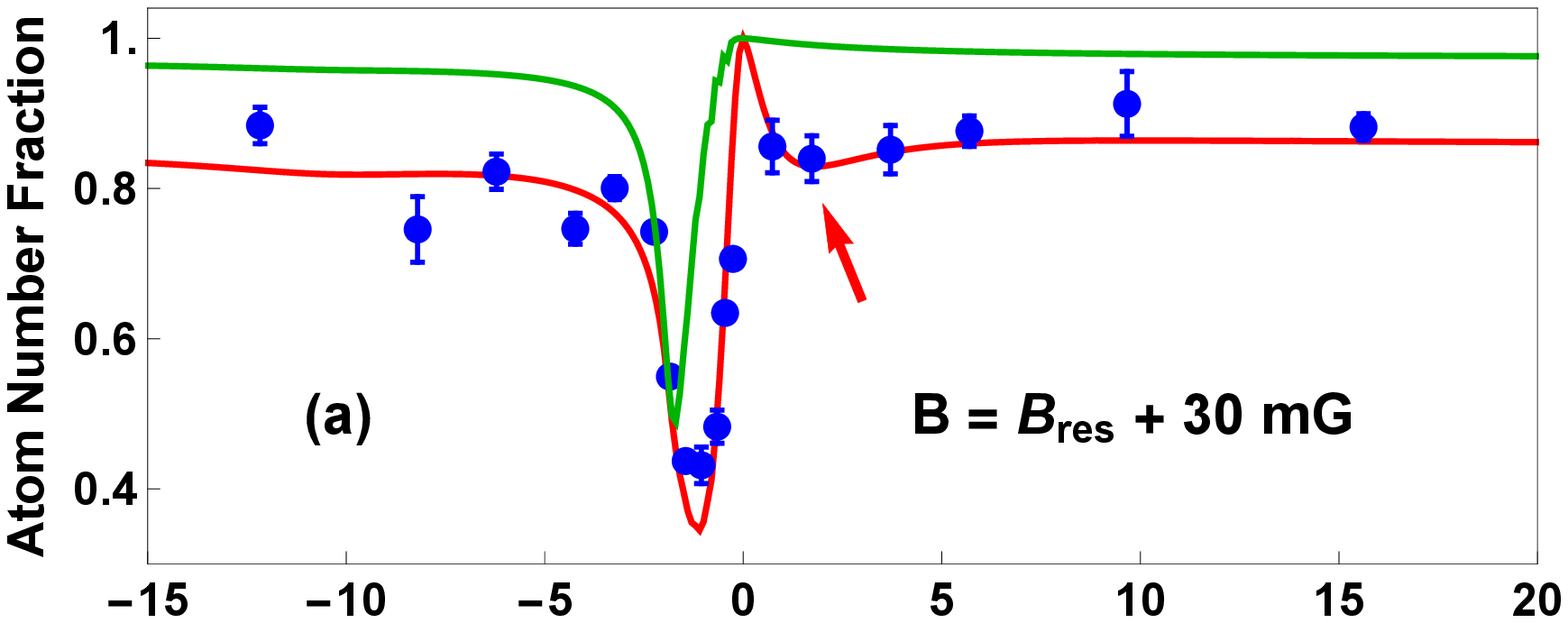}\\
	\vspace{0.1 in}
	\includegraphics[width= 3.4 in]{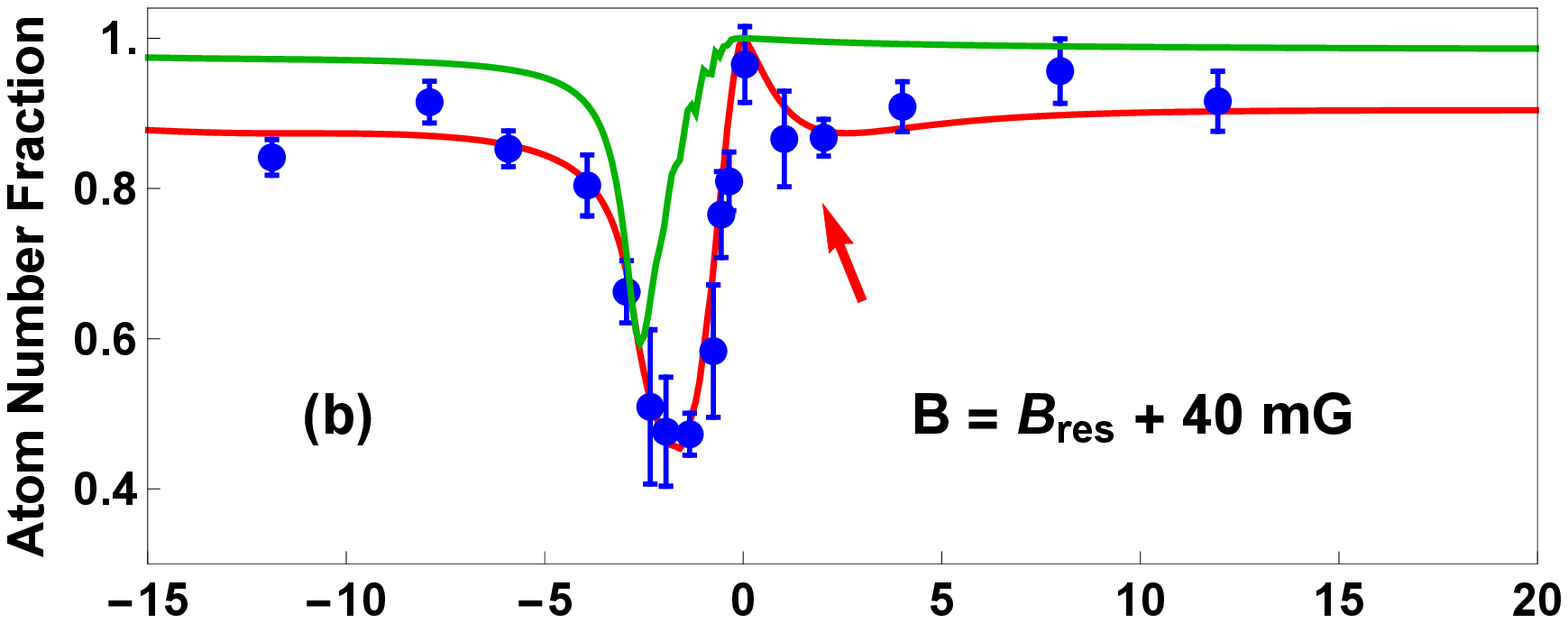}\\
	\vspace{0.1 in}
	\includegraphics[width = 3.4 in]{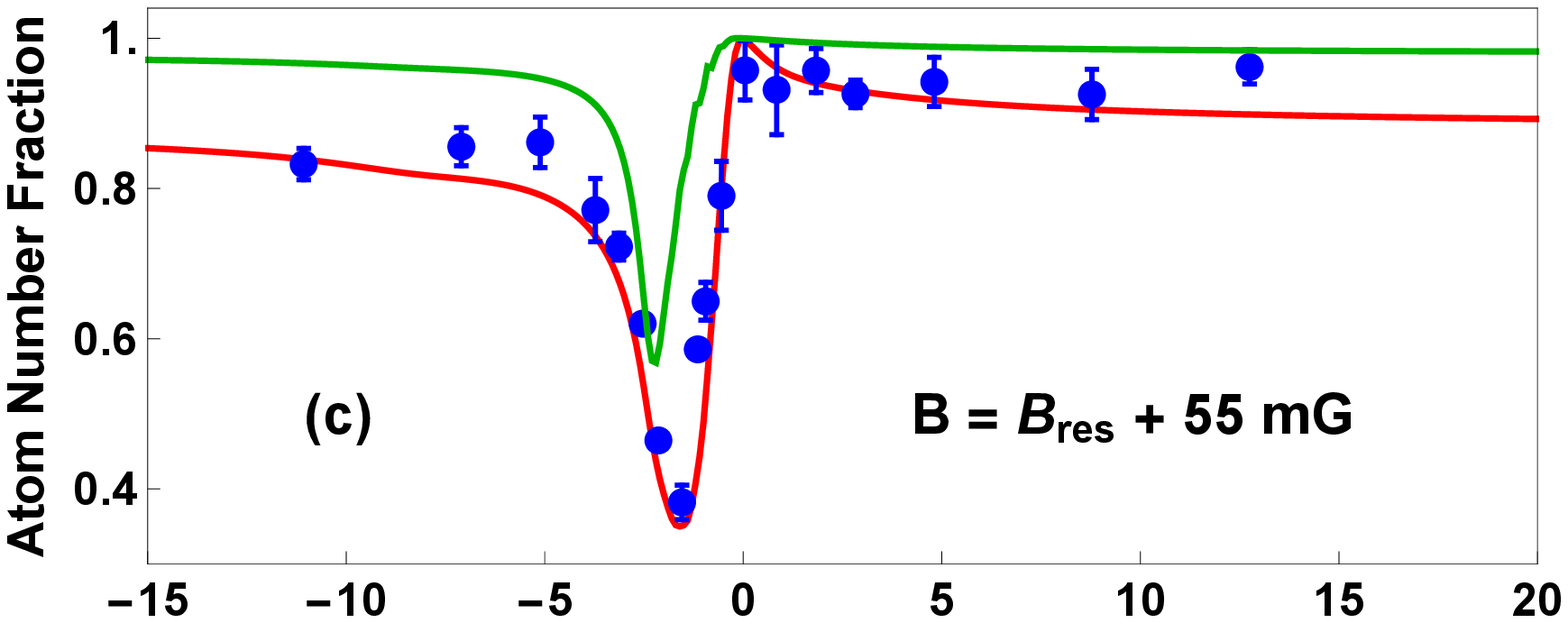}\\
	\vspace{0.1 in}
	\includegraphics[width = 3.4 in]{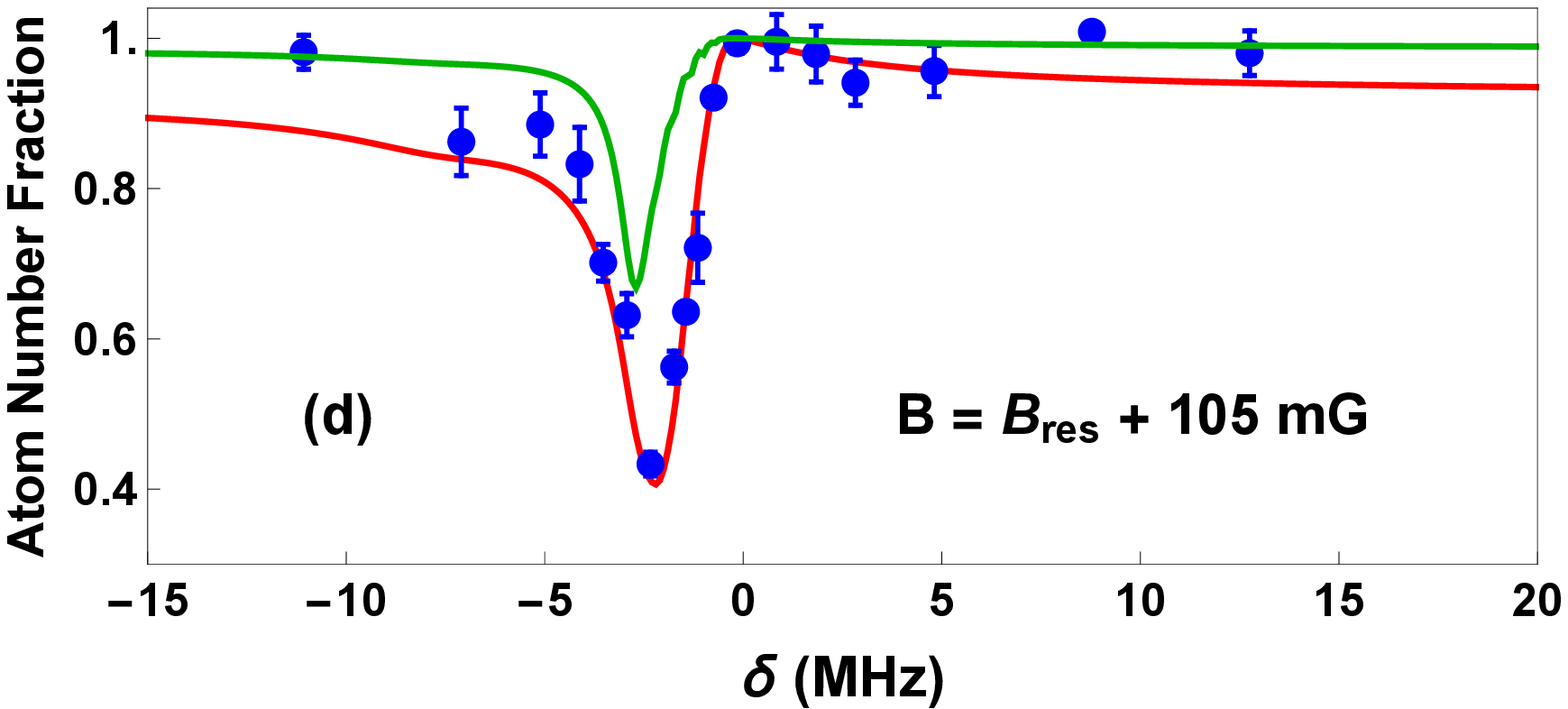}\\
	\caption{Two-photon atom loss spectra for $B>B_{res}$. Fraction of atoms remaining (blue dots) as a function of two-photon detuning $\delta = \Delta_2 - \Delta_1$. $\delta\equiv 0$ denotes the two-photon resonance. Red arrows denote secondary loss peaks (see text).  Solid curves: Predictions from $k$-averaged (red) and $k=0$ (green) theoretical model~\cite{SupportOnline,Range}.  Note that the data is shifted horizontally (see Fig.~\ref{ShiftvsB}) to align the measured minimum loss point with  theoretical minimum loss point at $\delta \equiv 0$. \label{opticalscatteringlengthBCS}}
\end{figure}

\begin{figure}[htb]
	\centering
	\includegraphics[width = 3.4 in]{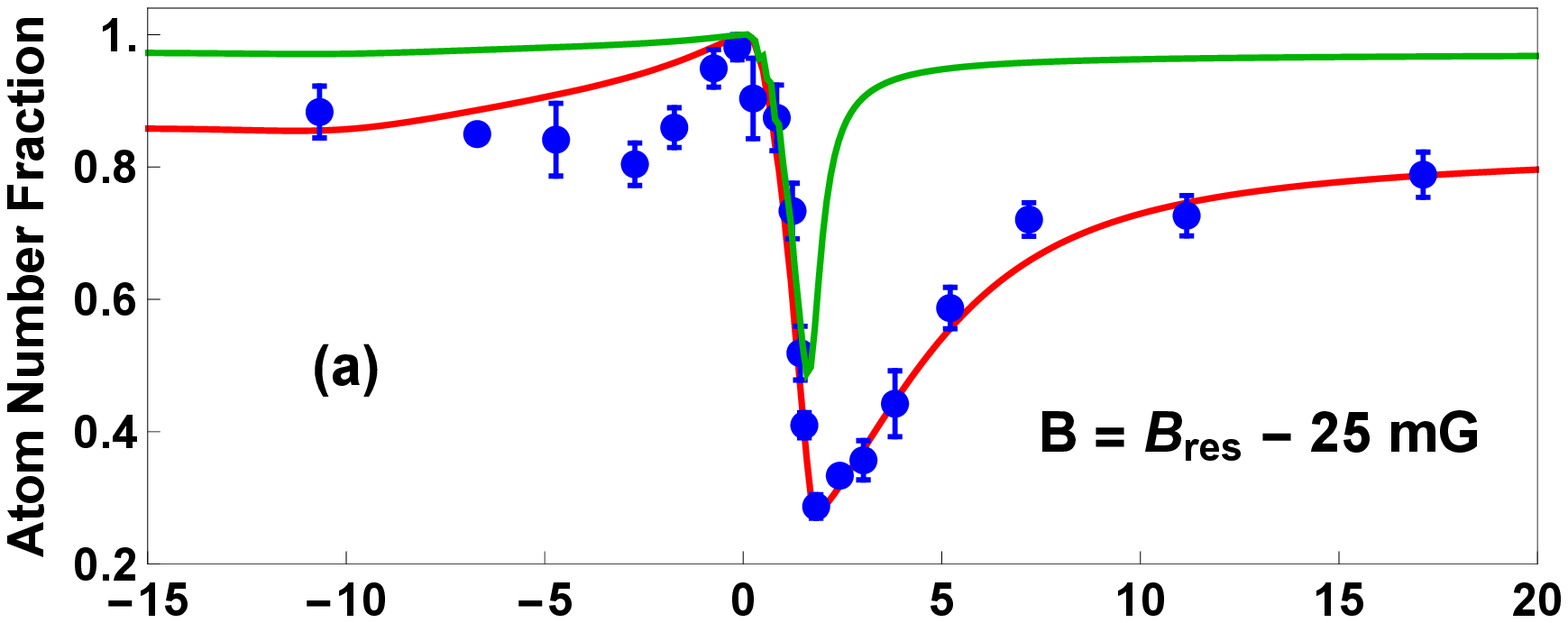}\\
	\vspace{0.1 in}
	\includegraphics[width= 3.4 in]{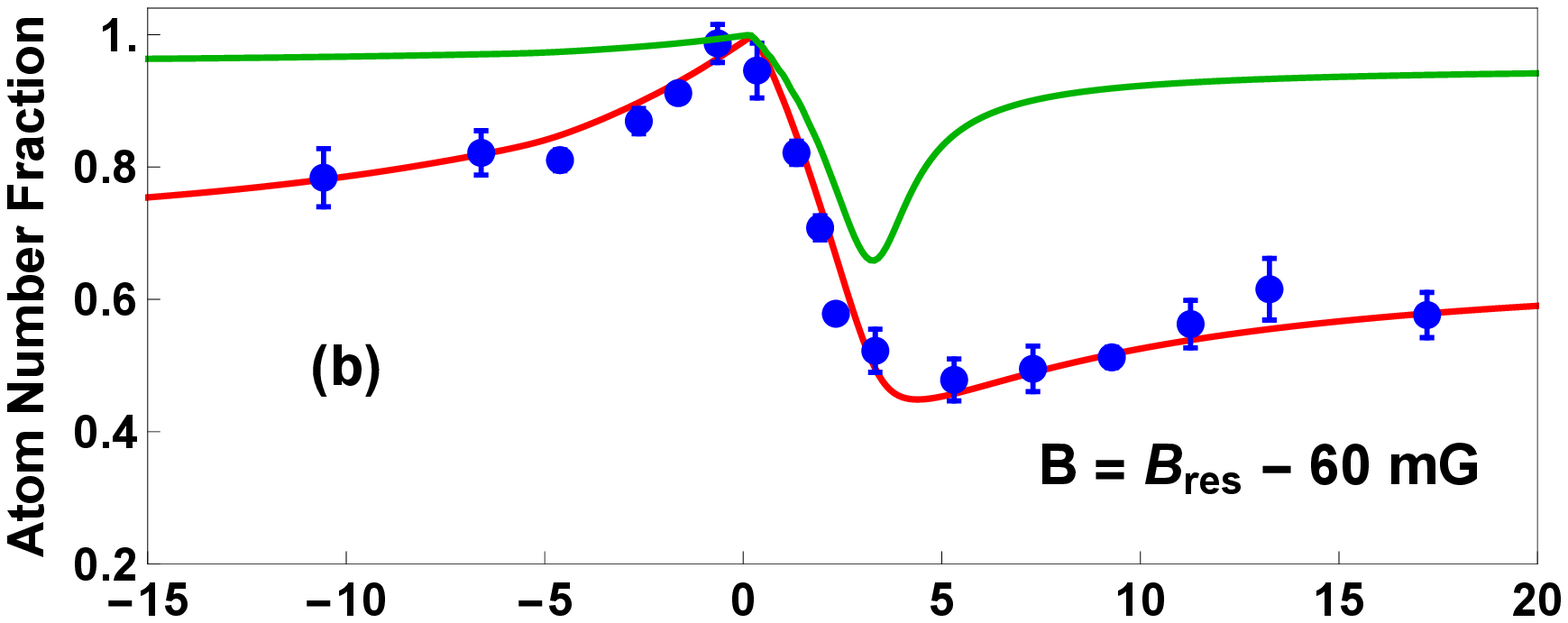}\\
	\vspace{0.1 in}
	\includegraphics[width = 3.4 in]{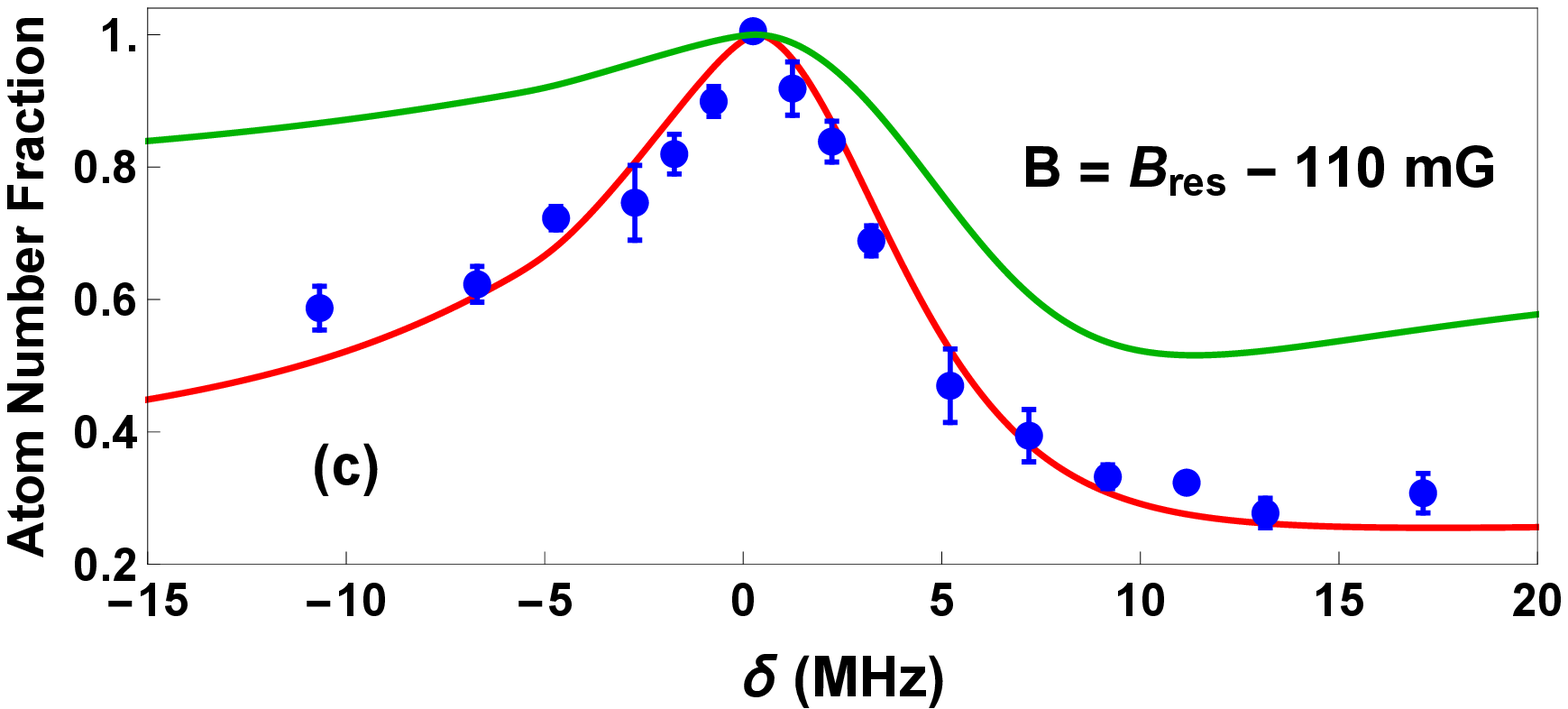}\\
	\caption{Two-photon atom loss spectra for $B<B_{res}$. Fraction of atoms remaining (blue dots) as a function of two-photon detuning $\delta = \Delta_2 - \Delta_1$, by varying $\nu_2$ and holding $\nu_1$ constant. $\delta\equiv 0$ denotes the two-photon resonance.   Solid curves: Predictions from $k$-averaged (red) and $k=0$ (green) theoretical model~\cite{SupportOnline,Range}. Note that the data is shifted horizontally (see Fig.~\ref{ShiftvsB}) to align the measured minimum loss point with  theoretical minimum loss point at $\delta \equiv 0$. \label{opticalscatteringlengthBEC}}
\end{figure}

\fref{opticalscatteringlengthBCS} and \fref{opticalscatteringlengthBEC} show widely different atom loss spectra for magnetic fields above resonance $B > B_{res}$ (BCS side) and below resonance $B < B_{res}$ (BEC side). Atom loss is measured versus two-photon detuning $\delta$,  by varying $\nu_2$ holding $\nu_1$ constant.  Atom loss arises from photoassociation of atoms in the triplet state $\ket{T,k}$,  Fig.~\ref{Horizontalimage}, where  hyperfine coupling $V_{HF}$ to $\ket{g_1}$ allows an optical transition to the excited singlet electronic state $\ket{e}$ and subsequent spontaneous emission.  Comparing \fref{opticalscatteringlengthBCS}a, 30 mG above resonance,  and \fref{opticalscatteringlengthBEC}a, 25 mG below resonance, we observe strikingly different spectral profiles, arising from the strongly momentum dependent scattering amplitude.

\begin{figure}[htb]
	\centering
	\includegraphics[width = 3.4 in]{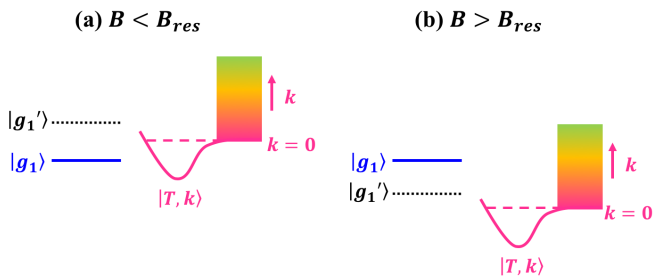}
	\caption{(a) For $B<B_{res}$, the threshold of the triplet continuum $\ket{T,k}$ is above the unshifted position of $\ket{g_1}$. Hence, no occupied momentum states are resonant with the unshifted state $\ket{g_1}$. However, the light-shifted state $\ket{g'_1}$ can be tuned into resonance with thermally occupied $k>0$ states. (b) For $B>B_{res}$, thermally occupied $k>0$ states of the continuum are resonant with the unshifted state $\ket{g_1}$, provided that the magnetic detuning $\frac{2\,\mu_B}{\hbar}\,(B-B_{res})$ is comparable to the thermal energy $\frac{\hbar^2 k^2}{m}$. For $B<B_{res}$ ($B>B_{res}$), maximum loss occurs for $\delta > 0$ ($\delta < 0$), where $\ket{g'_1}$ is tuned above (below) the unshifted state $\ket{g_1}$ to be resonant with the maximally occupied triplet continuum state $\ket{T,k}$.
		\label{MomentumInteractions}}
\end{figure}

To understand the effects of momentum-dependent interactions on the spectra in \fref{opticalscatteringlengthBCS} and \fref{opticalscatteringlengthBEC}, we compare the predictions of the continuum-dressed state  model~\cite{SupportOnline} for the $k$-averaged case (red solid line) to the zero momentum $k =0$ case (green solid line). The momentum-averaged model captures the fine features of the measured spectral profiles.  When the two-photon resonance condition is satisfied, i.e., $\delta \equiv 0$, atom loss is suppressed and the atom fraction $\simeq1$. As noted above, for $\delta =0$, the state $\ket{g_1}$ is also tuned to its original unshifted position. For both  $B > B_{res}$ (\fref{opticalscatteringlengthBCS}) and $B < B_{res}$ (\fref{opticalscatteringlengthBEC}), maximum loss in the spectra occurs for $\delta$ values greater than the prediction of the zero momentum $k = 0$ model (green curves). This illustrates that maximum loss occurs when state $\ket{g_1}$ is optically tuned to be degenerate with the maximally occupied state $\ket{T, k_{0}}$ and not with $\ket{T, 0}$. When atoms with $k > k_0$ and with $k<k_0$ are near the molecular bound state $\ket{g_1}$, \fref{MomentumInteractions}, the thermal momentum distribution averages positive and negative scattering amplitudes, resulting in a nearly zero mean field shift near resonance, as observed in the measurement of two-body interactions near a narrow Feshbach resonance~\cite{OHaraNarrowFB}.

\fref{opticalscatteringlengthBCS}a and \fref{opticalscatteringlengthBCS}b for $B > B_{res}$ show that the momentum-dependence of the scattering amplitude results in a second loss peak (red arrow) to the right of the minimum loss region (two-photon resonance $\delta = 0$), in agreement with the k-averaged theoretical model.
The spectral shapes are understood by considering the condition  for maximum loss. As shown by Eq.~S8 in the supplemental material of Ref.~\cite{ArunOptcontrol}, maximum loss occurs when
\begin{equation}
\Delta_1 + \frac{|\Omega_1|^2}{4 \,\big [ \frac{2\,\mu_B}{\hbar}\,(B-B_{res})-\frac{\hbar\,k^2}{m}\big]}+ \frac{|\Omega_2|^2}{4\,\delta} = 0.
\label{eq:1}
\end{equation}
Here, we assume for brevity that the frequencies corresponding to the magnetic detuning $\frac{2\,\mu_B}{\hbar}\,(B-B_{res})$ and kinetic energy $\hbar k^2/m$, are small compared to the  optical detunings $\Delta_1$ and $\Delta_2$.  For our experiments $\Delta_1 \approx + \,2\pi\times 19$ MHz. The second term is a one photon light shift of the singlet excited state, arising from off-resonant optical coupling to the full triplet k continuum, via the magnetic field dependent hyperfine mixing with the resonant singlet ground state~\cite{SupportOnline}. In our experiments, where $\Omega_1 \approx 2\pi\times 5.9$ MHz, and $|B-B_{res}| < 0.1$ G,  this optical shift term is large compared to the $\Delta_1$ term. Maximum loss therefore occurs when the one-photon optical shift is canceled by the two photon light shift given by the third term, where $\Omega_2\simeq 2\pi\times 26$ MHz in our experiments. When $B < B_{res}$, the $|\Omega_1|^2$ term in Eq.~\ref{eq:1} is negative for all $k$. Hence, the condition for maximum loss given by  Eq.~\ref{eq:1}  is satisfied \textit{only} when $\delta$ is positive, as we see in \fref{opticalscatteringlengthBEC}.

However, when $B > B_{res}$, loss peaks can occur for negative and positive $\delta$, as the $|\Omega_1|^2$ term in Eq.~\ref{eq:1} is  positive for $k<k_r$ and negative for $k> k_r$, where $\hbar k_r\equiv\sqrt{2\,\mu_B(B-B_{res})\,m}$ is the momentum of the triplet continuum state $\ket{T,k_r}$ that is degenerate with $\ket{g_1}$ in the absence of optical fields.
The primary loss peak occurs for $\delta < 0$, as the thermal atom population is larger for smaller $k<k_r$ (Maxwell-Boltzmann distribution). A second smaller loss peak occurs for $\delta > 0$, as $k>k_r$ is less populated, red arrows in \fref{opticalscatteringlengthBCS}a and \fref{opticalscatteringlengthBCS}b. Note the momentum-dependent model captures the observed loss peak for $\delta > 0$, while the $k = 0$ model (green curve) is completely flat. When the magnetic detuning $B-B_{res}$ is increased sufficiently, the thermal population of atoms with $k>k_r$ tends to zero and $k<k_r$ for all $k$. Therefore, the additional loss peak for $\delta > 0$ disappears with increasing magnetic field detuning, as shown in \fref{opticalscatteringlengthBCS}c and \fref{opticalscatteringlengthBCS}d.

For $B < B_{res}$, \fref{opticalscatteringlengthBEC},  the unshifted molecular bound state $\ket{g_1}$ lies below the triplet continuum and cannot resonantly couple with any $k$ states in the absence of optical fields. However, $\ket{g_1}$ can be optically tuned to be resonant with non-zero $k$ states by increasing $\delta$. This appears as a long tail on the right side of the maximum loss peak in the spectra, where increasing $\delta$ scans $\ket{g_1'}$  through the thermal distribution of $k$ states, \fref{MomentumInteractions}.

\begin{figure}[htb]
	\centering
	\includegraphics[width = 3.0 in]{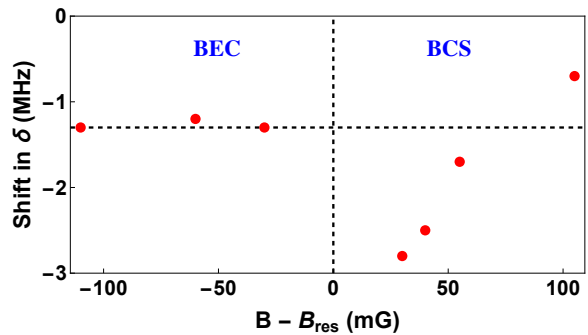}
	\caption{Spectral shift of two-photon loss spectra (shown in \fref{opticalscatteringlengthBCS} and \fref{opticalscatteringlengthBEC}) as a function of magnetic field $B$ near the narrow Feshbach resonance at $B_{res} = 543.27$ G in $^6$Li. The horizontal dashed line shows a background shift of $\simeq -1.3$ MHz. Vertical dashed line; $B = B_{res}$.
	\label{ShiftvsB}}
\end{figure}

The shapes and magnitudes of the complex spectra, which determine the momentum dependence of the scattering amplitude, are very well fit by our continuum dressed state model, using temperatures and Rabi frequencies that are within 20\% of the values estimated from the beam size and cloud profile~\cite{Range}. However, we observe overall frequency shifts between the measured and predicted two-photon spectra shown in \fref{opticalscatteringlengthBCS} and \fref{opticalscatteringlengthBEC}, which are unexplained. In the figures, the data has been frequency shifted (see \fref{ShiftvsB}) to align the measured minimum loss point with the theoretical minimum loss point at $\delta=0$. For $B<B_{res}$, we observe that the required shift of the data, $\simeq -1.3$ MHz,  is nearly independent of magnetic field, while for $B>B_{res}$, we observe a strong  magnetic field dependence. The magnetic field independent shift may arise in part from a systematic error or from an anomalous background redshift of the excited state~\cite{JulienneAnomalousShift,WalravenOptTuning}, which would affect our $\Delta_1$ frequency calibration and hence the absolute value of the two photon detuning $\delta$. Such an anomalous shift was observed in the photoassociation experiments in $^7$Li~\cite{RandyIntensityShift,RandyAnomalousShift}, resulting in an overall spectral shift of the single-field atom loss spectra. An additional intensity-dependent asymmetric shift has been observed near a Feshbach resonance in $^7$Li~\cite{RandyBroadAnomalousShift,TheoryAnomalousShiftBroad}. As noted above, the one-photon optical shift term $\propto|\Omega_1|^2$ in Eq.~\ref{eq:1}  arises from the Feshbach resonance induced optical coupling of the triplet continuum to the excited state. This term appears to explain the asymmetric shift observed in Ref.~\cite{RandyBroadAnomalousShift}. However, the  frequency shifts observed in our experiments for the two-photon detuning are not explained and require further study.

In summary,  we have studied momentum-dependent interactions for a narrow Feshbach resonance, by optically tuning the closed channel molecular bound state near the open channel continuum threshold. Using a closed-channel EIT method as an optical vernier, we observe that the momentum dependence of the two-body scattering amplitude strongly modifies the two-photon atom-loss spectra, providing new insights into energy-dependent Feshbach resonances. Variants of our two-field optical vernier method will have important applications, for example, to create paired states with nonzero center of mass momentum~\cite{OptcontrolCOM, FuldeFerrell}, as suggested recently.

This research is supported by the Physics Divisions of the Air Force Office of Scientific Research  and the National Science Foundation. Additional support is provided by the Army Research Office  and the Division of Materials Science and Engineering, the Office of Basic Energy Sciences, Office of Science, U.S. Department of Energy.


\begin{thebibliography}{32}
\bibitem[{Ran()}]{Range}
\bibinfo{note}{The best fits to the data taken over several months are obtained
  for Rabi frequencies in the range $\Omega_2=26.0(5.0)$ MHz,
  \mbox{$\Omega_1=5.9(0.6)$ MHz}, and temperatures in the range
  $1.8(0.3)\,\mu$K. These parameters are within 20 percent of the measured
  values determined from the beam and cloud profiles, except for Fig. 3b, where
  the best fit is obtained for a higher temperature of $3.0\,\mu$K. The range
  arises from day to day changes in the beam alignment and power.}
\end{thebibliography}

\end{document}